\documentclass[prb,aps,twocolumn,amsmath,amssymb,floatfix,superscriptaddress]{revtex4}

\usepackage[dvips]{graphics}
\usepackage{color}
\usepackage{soul}
\usepackage{comment}
\usepackage{mathrsfs}
\usepackage[colorlinks=true, citecolor=blue, urlcolor=blue]{hyperref}

\date{\today}

\begin{document}

\title{Flux-Driven Circular Current in a Non-Hermitian Dimerized Aharonov-Bohm Ring: Impact of Physical Gain and Loss}
	
\author{Souvik Roy}

\email{souvikroy138@gmail.com}

\affiliation{School of Physical Sciences, National Institute of Science Education and Research Bhubaneswar, Jatni-752 050, India}

\affiliation{Homi Bhabha National Institute, Training School Complex, Anushaktinagar, Mumbai-400 094, India}
	
\author{Santanu K. Maiti}

\email{santanu.maiti@isical.ac.in}
	
\affiliation{Physics and Applied Mathematics Unit, Indian Statistical Institute, 203 Barrackpore Trunk Road, Kolkata-700 108, India}
	
\begin{abstract}

In the present theoretical work, we numerically explore magnetic response of a tight-binding dimerized ring subjected to Aharonov-Bohm (AB) 
flux and environmental interactions.
Specifically, we introduce an imaginary site potential on the odd lattice sites to represent physical gain and loss, while the even lattice
sites remain unperturbed. We investigate the induced current resulting from the AB flux in both real and imaginary eigenspaces, aiming to
enhance this current significantly by adjusting the gain/loss parameter ($d$). Our analysis focuses on how exceptional points in the real 
and imaginary eigenenergy spaces contribute to notable increases in current at specific $d$ values, and the emergence of purely real current
when the imaginary current vanishes. We discuss how the dual behavior of energy spectrum (real and imaginary), converging to and diverging 
from zero energy, affects the enhancement of the current. Additionally, we study the interplay between the correlations of dimerized 
hopping integrals and the gain-loss parameter, which affects the current and highlights key features associated with these physical 
parameters. Furthermore, we consider how system size impacts our findings. Our study may reveal unconventional characteristics in various 
loop configurations, potentially paving the way for new research directions.

\end{abstract}
 
\maketitle

\section{Introduction}

Recent studies have focused intensely on various non-Hermitian (NH) systems, particularly those that transform into parity-time (\emph{PT}) symmetric configurations upon accounting for environmental interactions~\cite{r1,r2,r3,r4,r5}. Such systems have garnered significant attention across research domains. Non-Hermiticity in tight-binding systems can arise from either non-reciprocal hopping integrals or complex site potentials or from both. Depending on the sign of the complex potential, these systems may exhibit either gain or loss, crucial for maintaining \emph{PT} symmetry through balanced distributions. Under specific parameter regimes, \emph{PT} symmetric NH systems~\cite{r6,r7,r8,r9,r10} can feature a real energy spectrum and display intriguing characteristics. Notably, exceptional points and non-orthogonal eigenmodes have been identified in optical NH systems, where proximity to exceptional points has been associated with phenomena like heightened sensing, sensitivity, diffusive coherent transport, topological states, and the skin effect as documented in literature. Such NH and \emph{PT} symmetric systems~\cite{r11,r12,r13,r14,r15,r16,r16a} have also been extensively explored in electronics, sound propagation, electromagnetic wave propagation in media with complex refractive indices, and aspects of atomic and nuclear physics. Theoretical and experimental investigations abound, probing into the diverse physics inherent in these systems.
In addition to these, localization phenomena in closed systems and quantum transport~\cite{r17,r18,r19,r20,r21,r22,r23} in open quantum systems have been investigated in various NH systems. In the analysis of quantum transport, exceptional points play a crucial role in probing the behavior of transmission function. The influence of physical gain and loss~\cite{r24,r25,r26,r27,r28,r29,r30,r31,r32,r33,r34,r35,r36,r37,r38,r39,r40,r41,r42,r43,r44,r45,r46,r47,r48,r49,r50,r51,r52,r53} is also significant in the context of extended-to-localized phase transitions~\cite{r54,r55,r56,r57,r58,r59,r60,r61,r62,r63,r64,r65,r66,r67}. Exceptional points amplify the generation of circular currents and consequently induce a magnetic field in open quantum systems. Similarly, the behavior of circular currents in dimerized closed-loop geometries under AB flux can be examined, where odd lattice sites experience physical gain and loss, while even lattice sites remain perfect. 
\begin{figure}[ht]
\hskip 1.0in
\noindent
{\centering\resizebox*{9cm}{5.5cm}{\includegraphics{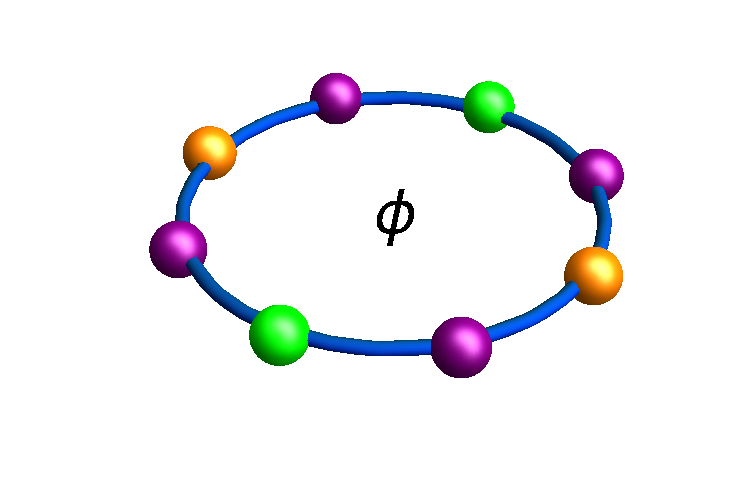}\par}}
\caption{(Color online). This diagram illustrates a one-dimensional ring system subjected to environmental interactions and threaded by an AB flux $\phi$. The purple colored spheres represent the unperturbed lattice sites (viz, free from NH interaction), while the orange and green colored spheres denote the lattice sites in the presence of gain and loss, respectively.}
\label{fig:f1}
\end{figure}
To the best of our knowledge, such an investigation has not been documented in the literature. Exploring the behavior of circular currents under AB flux in such NH systems may provide several crucial facts of quantum loop geometries. In the present work, we aim to explore that in a systematic way. 
\begin{figure*}[ht]
\noindent
{\centering\resizebox*{15.0cm}{12.5cm}{\includegraphics{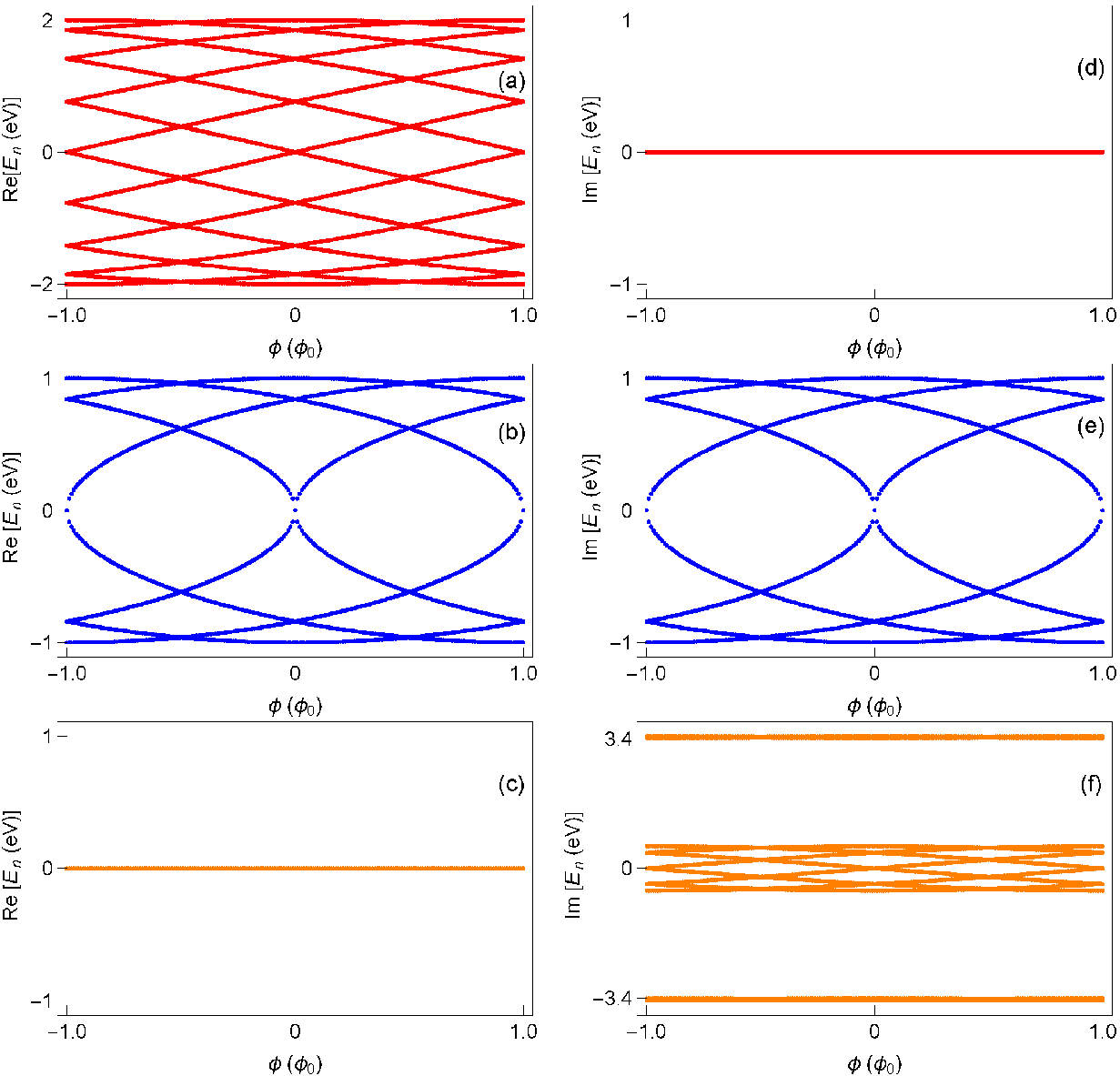}\par}}
\caption{(Color online). Real (left column) and imaginary (right column) energy eigenvalues as a function of magnetic flux $\phi$ for a $16$-site ring in the dimerized-free case ($t_1=t_2$). The distinct colors are used to differentiate the values of the NH gain and loss parameter, $d$: the red, blue, and orange lines correspond to $d=0$, $2$, and $4$, respectively.}
	\label{fig:f2}
\end{figure*}
\begin{figure*}[ht]
\noindent
{\centering\resizebox*{17.0cm}{10cm}{\includegraphics{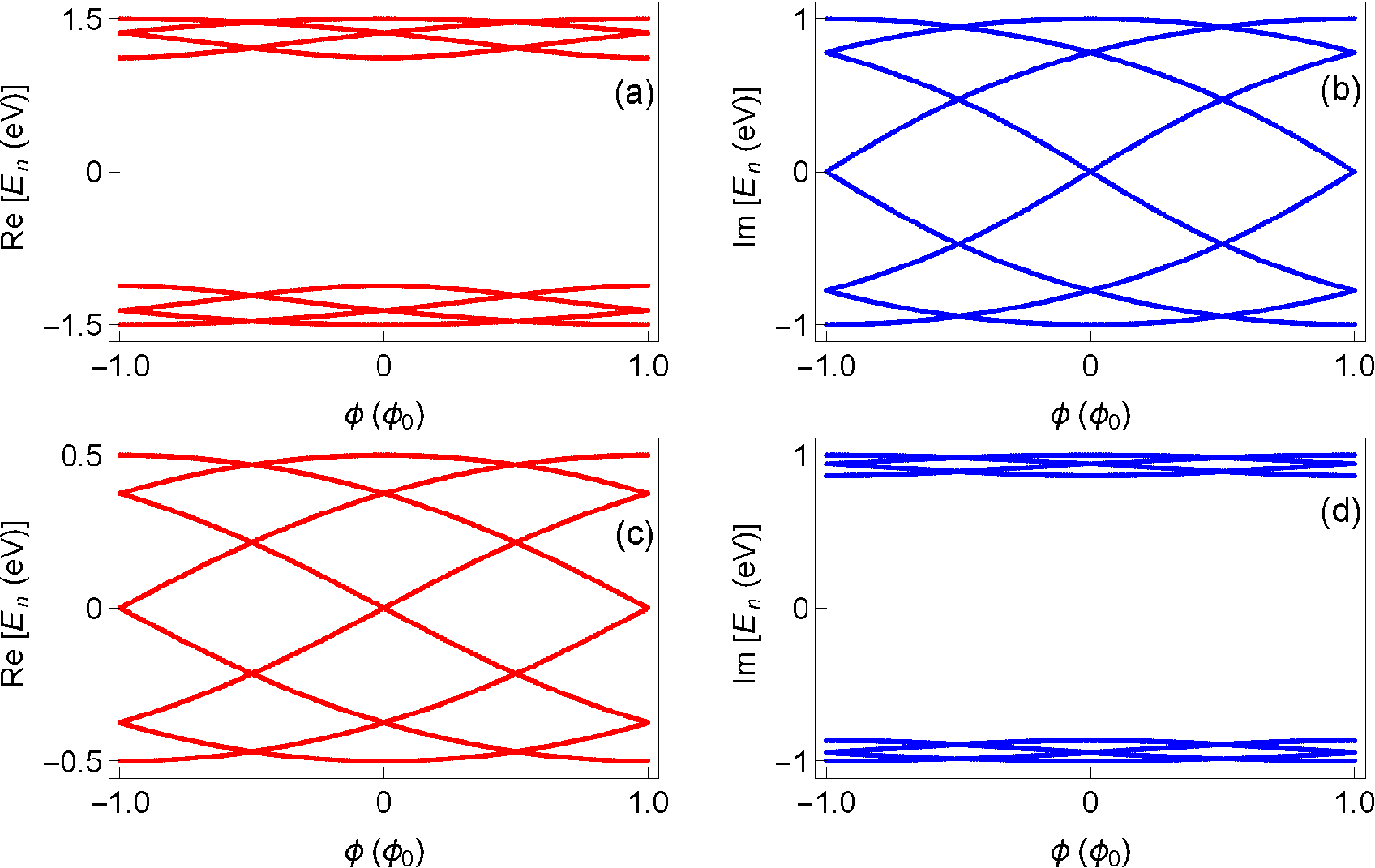}\par}}
\caption{(Color online). Variation of the real (left column) and imaginary (right column) energy eigenvalues as a function of flux $\phi$
for a $16$-site ring with $d=2$, in the hopping dimerized case. The top and bottom rows correspond to the conditions $t_1>t_2$ and $t_1<t_2$,
respectively.}
\label{fig:f3}
\end{figure*}
\begin{figure*}[ht]
\noindent
{\centering\resizebox*{14.0cm}{11.0cm}{\includegraphics{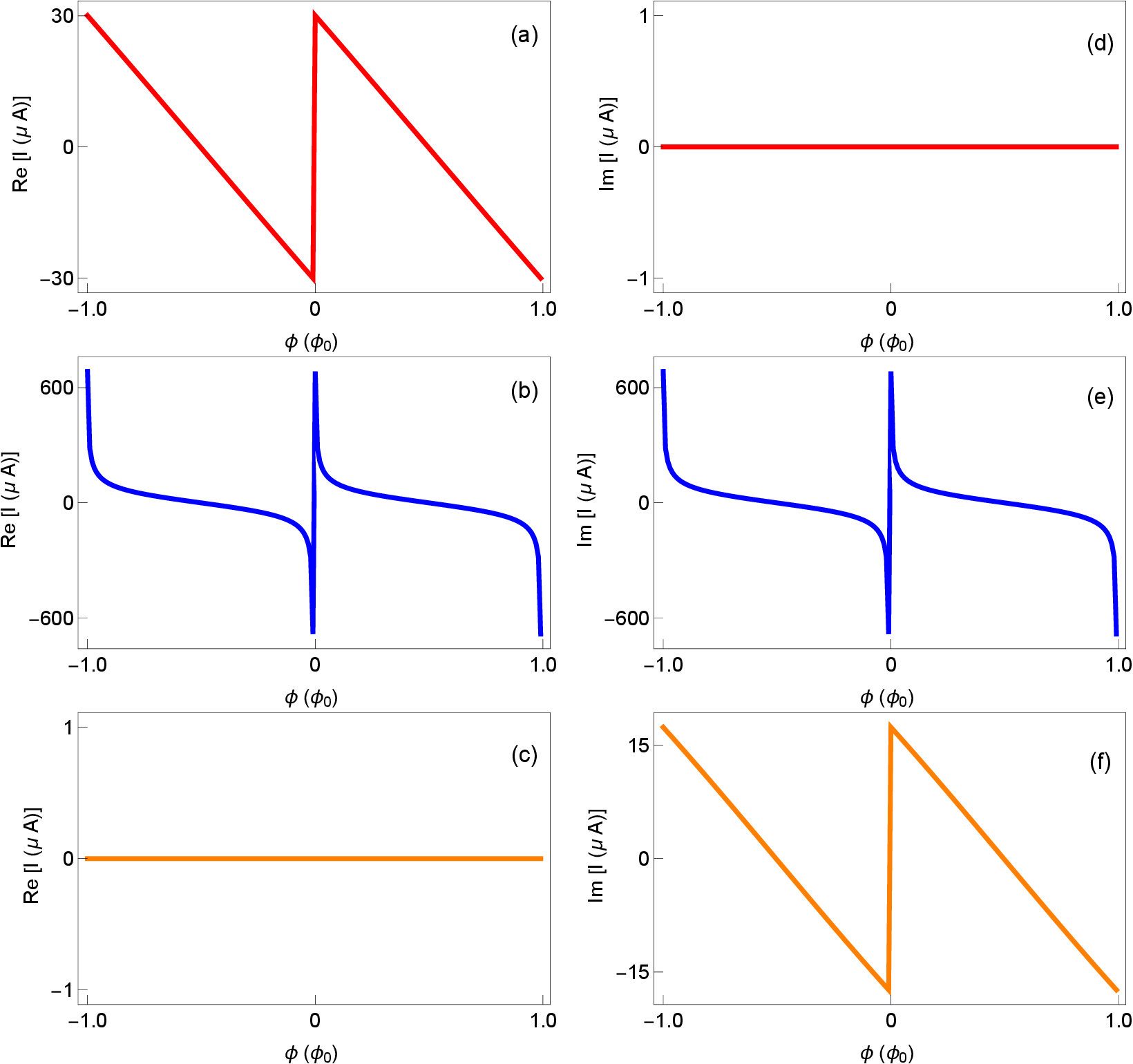}\par}}
\caption{(Color online). Real (left column) and imaginary (right column) circular currents as a function of $\phi$ in the ring system with 
identical set of parameters as mentioned in Fig.~\ref{fig:f2} ($N=16$, $t_1=t_2$). The upper, middle, and the lower rows correspond to 
$d=0$, $2$, and $4$, respectively.}
\label{fig:f4}
\end{figure*}
\begin{figure*}[ht]
	\noindent
{\centering\resizebox*{16.0cm}{12.0cm}{\includegraphics{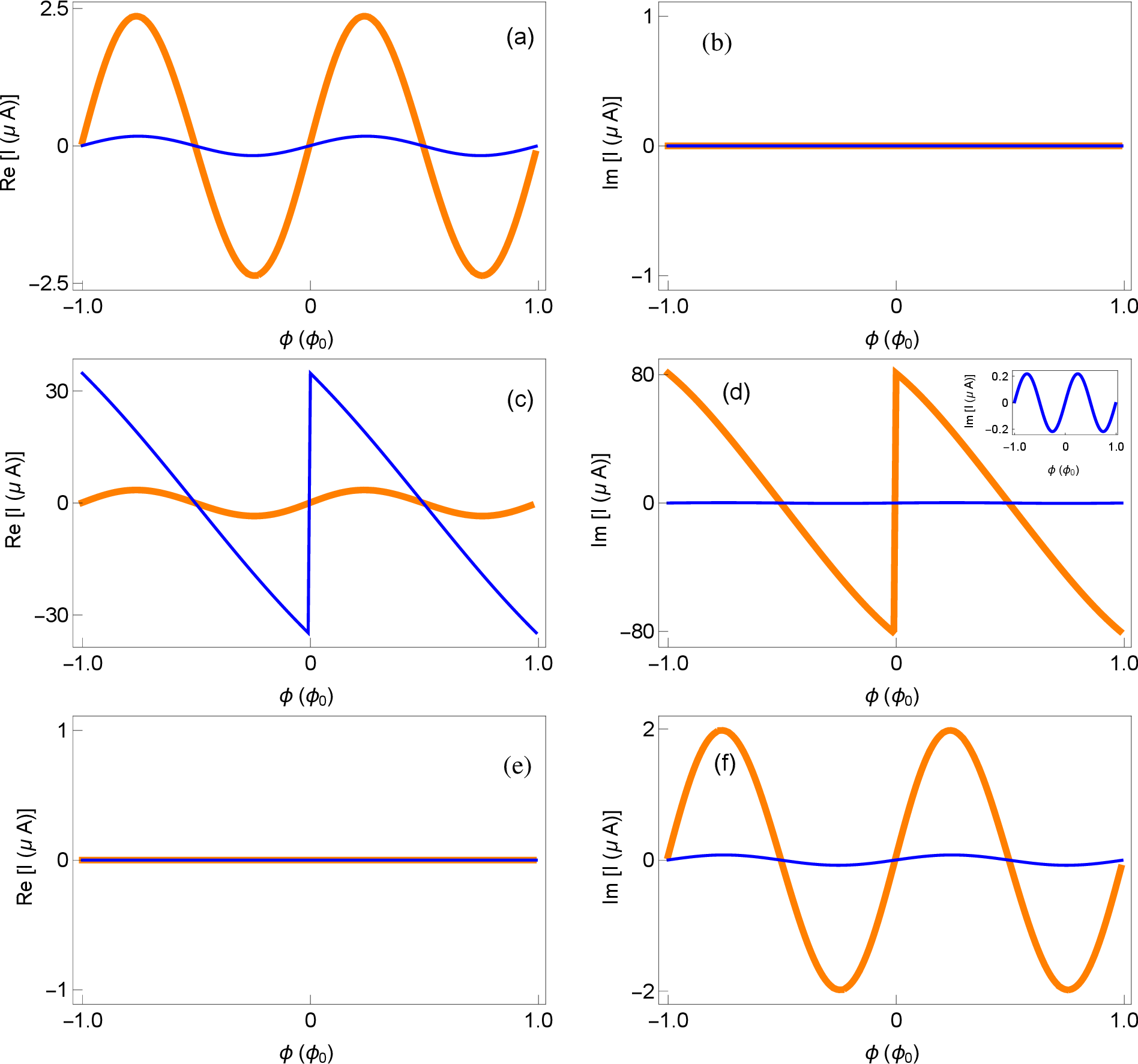}\par}}
\caption{(Color online). Real (left column) and imaginary (right column) circular currents in a $16$-site ring under the dimerized condition, where the blue and orange curves correspond to $t_1<t_2$ and $t_1>t_2$, respectively. We use a thicker orange curve compared to the blue one 
to ensure better visibility in black and white print format. The first, second, and the third rows are for $d=0$, $2$, and $4$, respectively.}
\label{fig:f5}
\end{figure*}
\begin{figure*}[ht]
	\hskip 1.2in
	\noindent
{\centering\resizebox*{16.0cm}{10.0cm}{\includegraphics{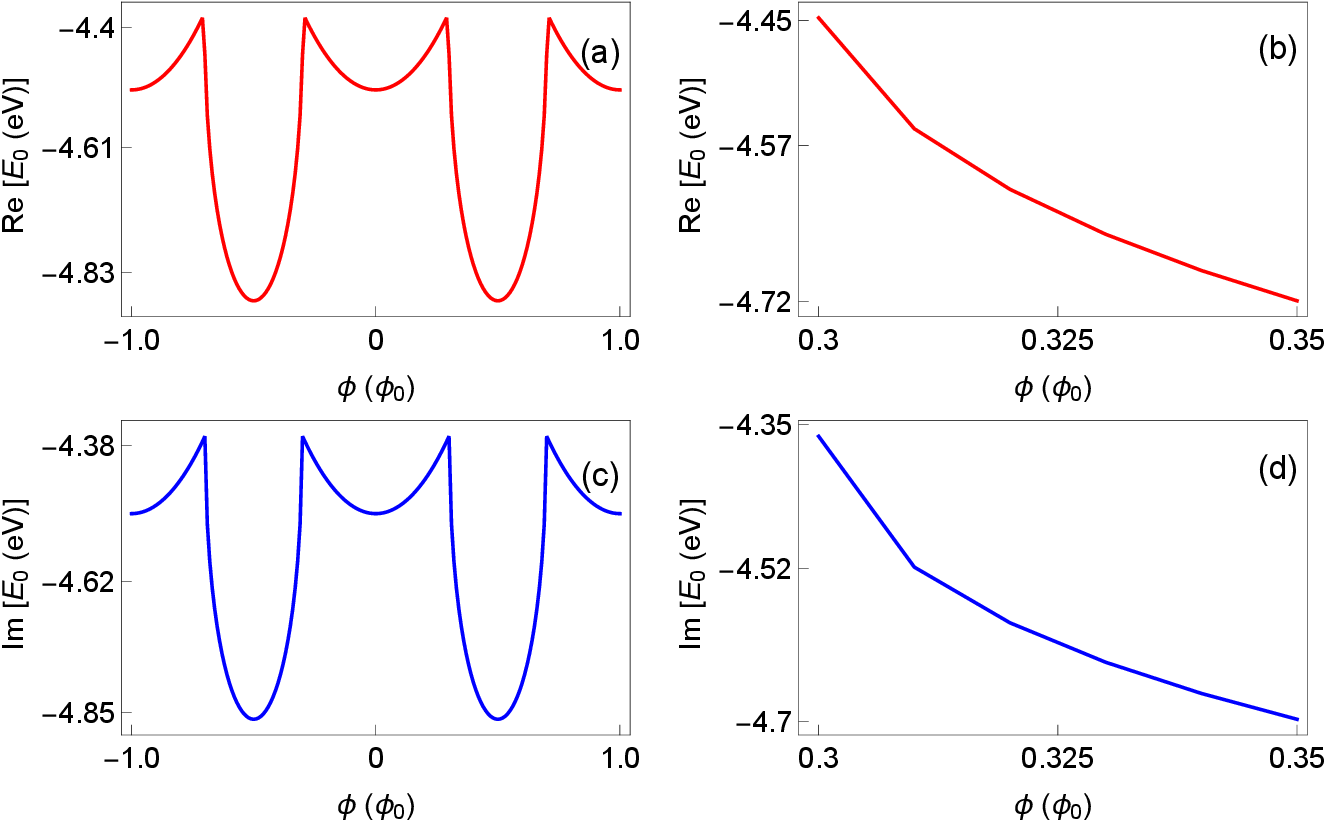}\par}}
\caption{(Color online). Left column: Variation of the real (red curve) and imaginary (blue curve) ground state energies as a function of 
magnetic flux, at the first exceptional points where the real and imaginary currents show enhancements. For the real case $d$ is $2.22$, while
it is $1.75$ for the imaginary case. These are the first exceptional points in the two different energy spaces. Right column: For a better 
viewing of the slopes, we select a small flux window and re-plot the ground state energies. These results are shown for a $16$-site ring, 
considering $t_1=t_2$.} 
	\label{fig:f6}
\end{figure*}
\begin{figure*}[ht]
	\noindent
{\centering\resizebox*{18.0cm}{14.0cm}{\includegraphics{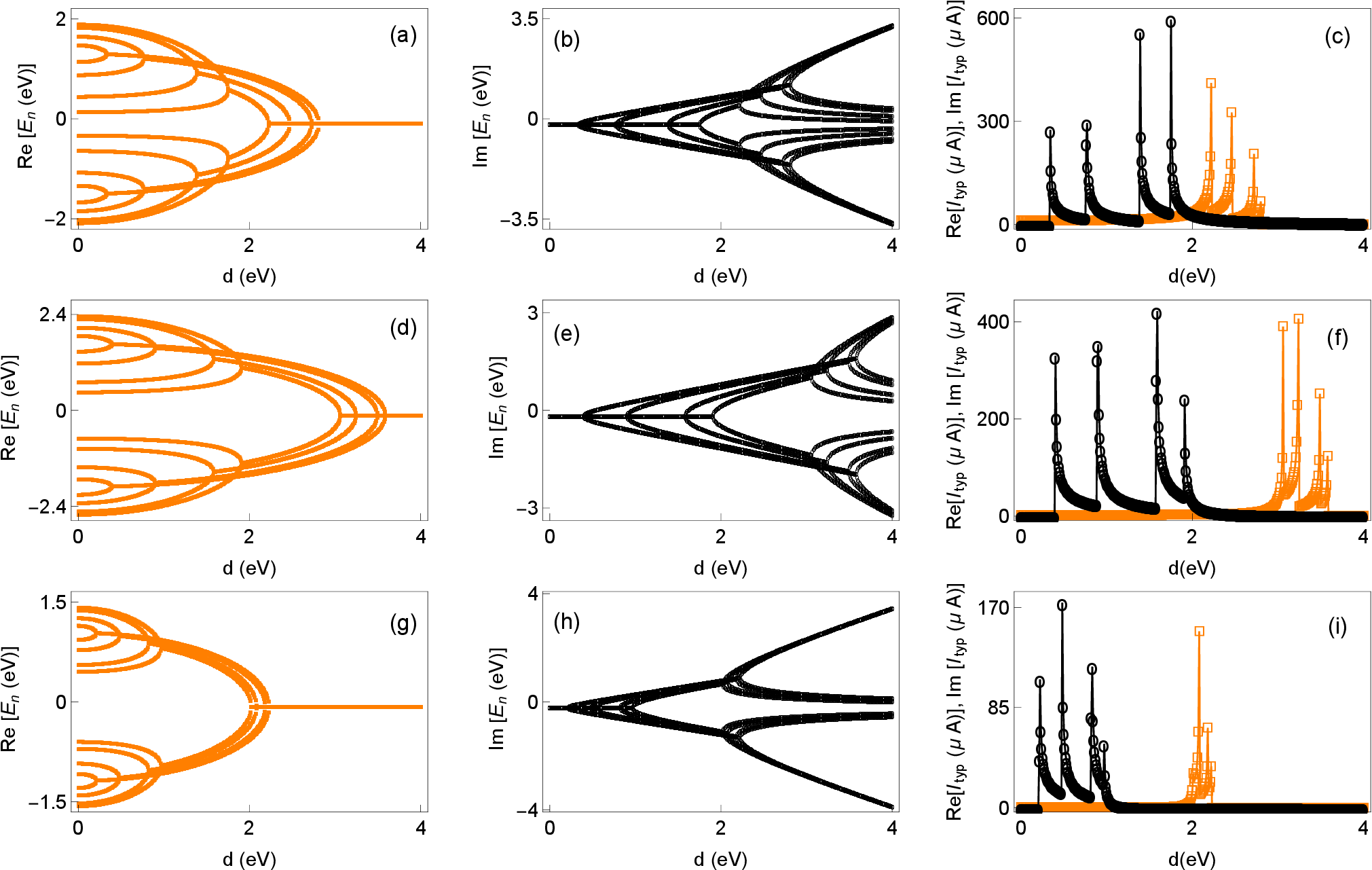}\par}}
\caption{(Color online). Characteristic features of the real energy eigenvalues (left column), imaginary energy eigenvalues (middle column), 
and the real and imaginary currents (right column) as a function of the gain-loss parameter $d$ under three different conditions of the 
hopping integrals: $t_1=t_2$ (top row), $t_1>t_2$ (middle row), and $t_1<t_2$ (bottom row). In the current spectra, the real and imaginary 
components are described by the orange and black lines, respectively. Two different line styles of the currents are used for a clear 
distinction in the black and white print format. The results are shown for the ring with $N=16$, considering magnetic flux $\phi=0.3$.}
\label{fig:f7}
\end{figure*}
\begin{figure*}[ht]
\noindent
{\centering\resizebox*{17cm}{13cm}{\includegraphics{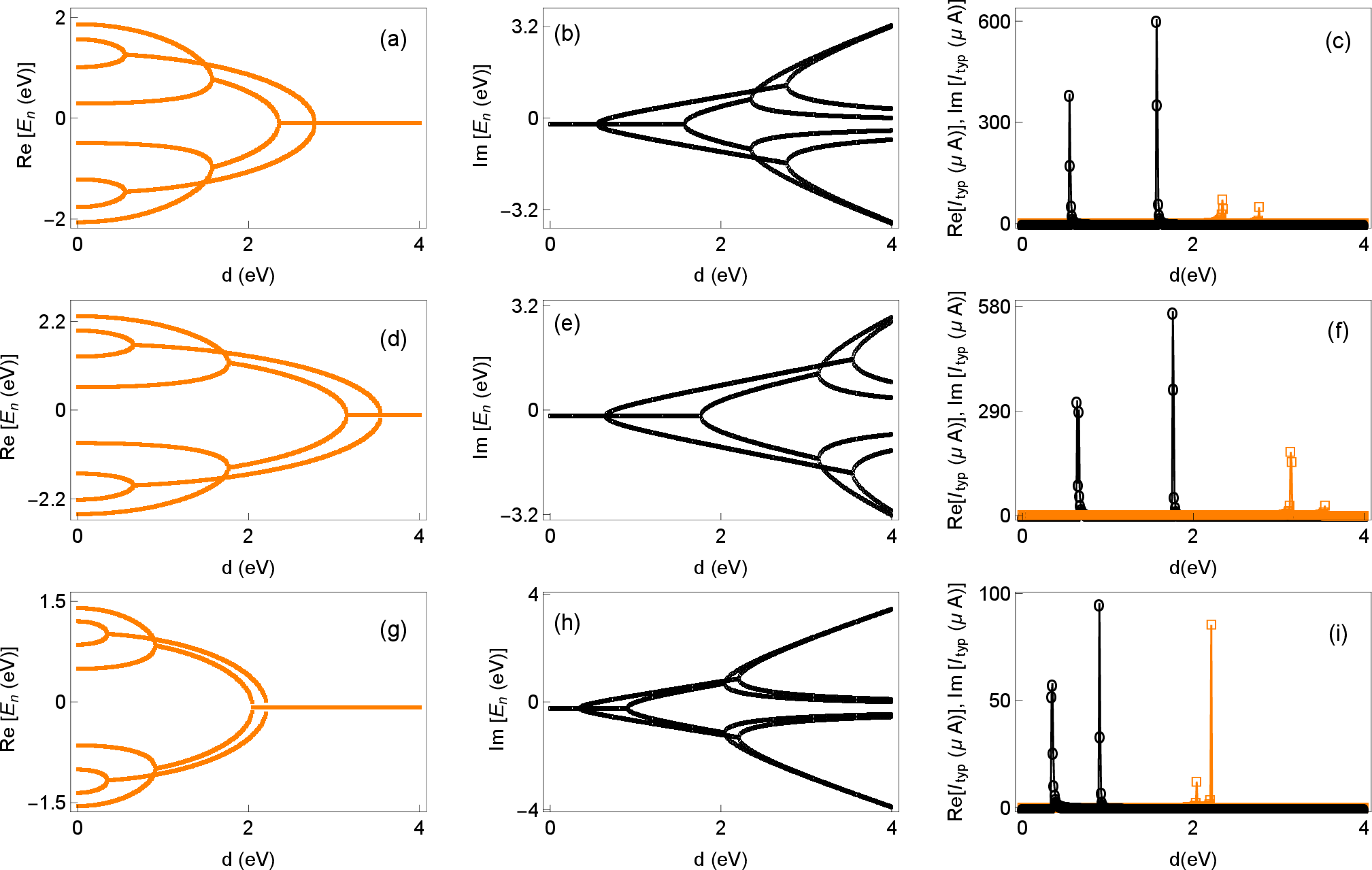}\par}}
\caption{(Color online). Same set of plots as shown in Fig.~\ref{fig:f7}, with magnetic flux $\phi=0.5$.}
\label{fig:f8}
\end{figure*}
\begin{figure*}[ht]
	\hskip 1.2in
	\noindent
	{\centering\resizebox*{16.0cm}{12.0cm}{\includegraphics{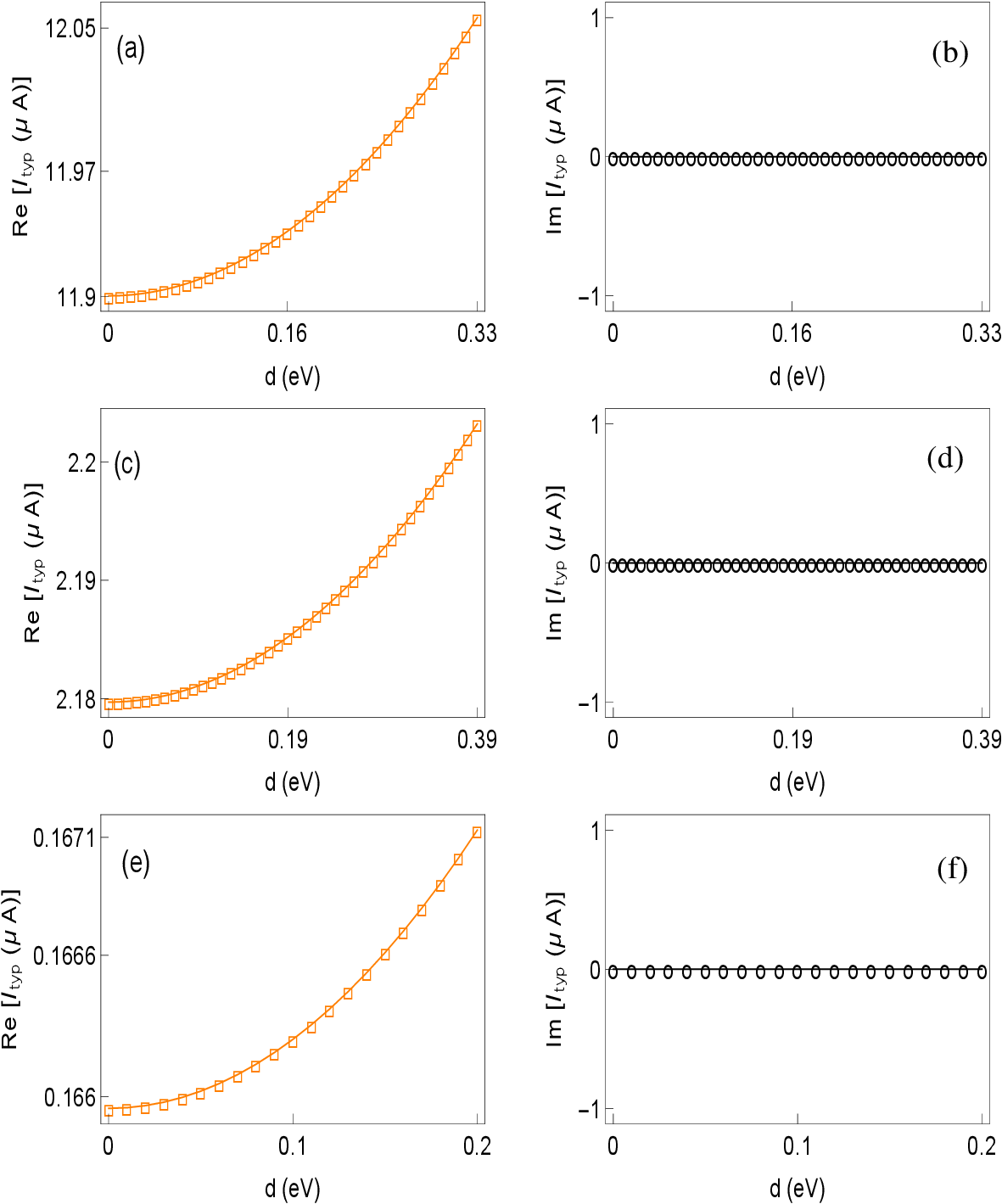}\par}}
\caption{(Color online). Variation of the real current (left column) up to certain value of $d$ for which the imaginary current (right column)
remains zero. The results are shown for three different conditions of hopping integrals, like earlier, where the top, middle, and the bottom 
rows correspond to $t_1=t_2$, $t_1>t_2$, and $t_1<t_2$, respectively. Here we set $N=16$ and $\phi=0.3$.}
\label{fig:f9}
\end{figure*}
\begin{figure*}[ht]
\hskip 1.2in
\noindent
{\centering\resizebox*{17.0cm}{11.0cm}{\includegraphics{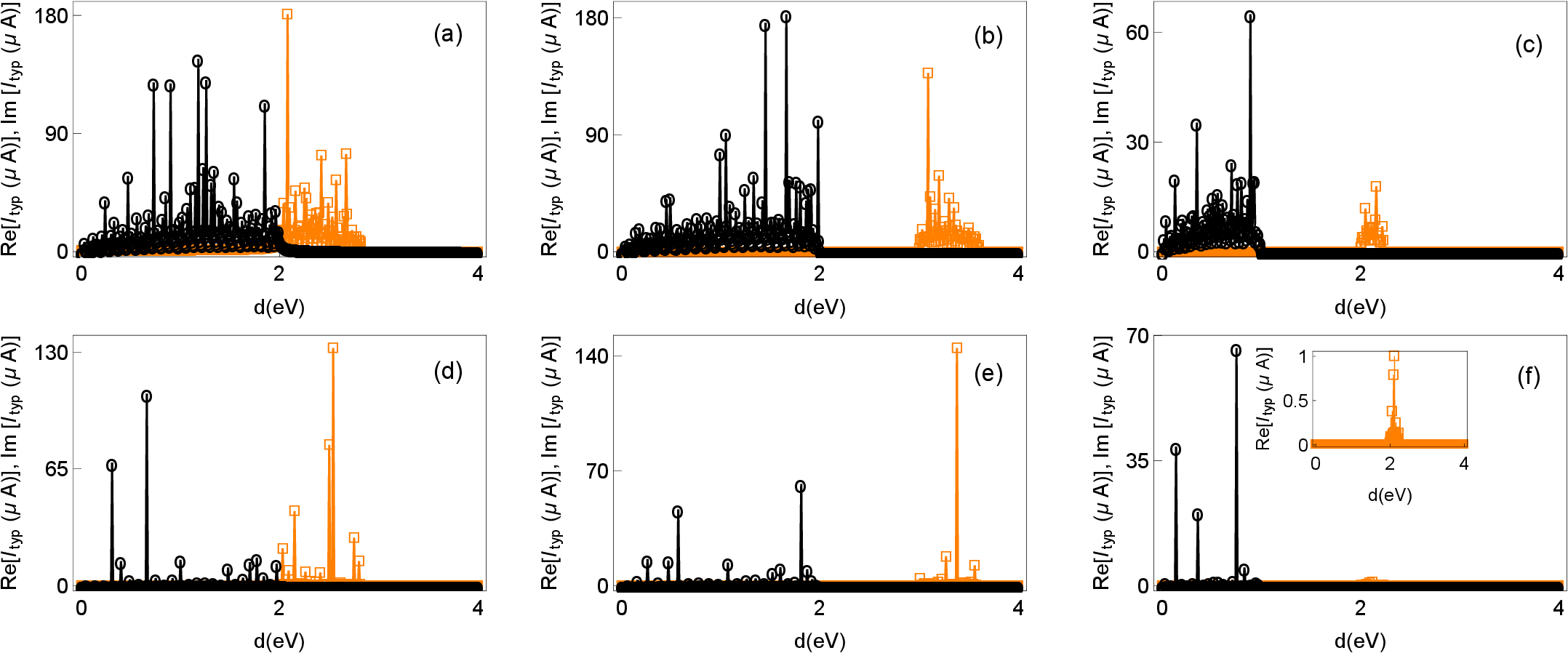}\par}}
\caption{(Color online). Effect of ring size $N$ on currents is shown. Each sub-figure contains the real (orange curve with empty squares) 
and the imaginary (black curve with empty circles) currents as a function of $d$, where the left, middle, and the right columns are for
$t_1=t_2$, $t_1>t_2$, and $t_1<t_2$, respectively. The results are computed for $N=200$. The upper and lower rows correspond to $\phi=0.3$
and $0.5$, respectively.} 
\label{fig:f10}
\end{figure*}

According to B\"{u}ttiker, Imry, and Landauer~\cite{r67,r68,r69,r70}, if the current is originated in a ring in presence of AB flux, it will 
remain in the ring even after removal of flux~\cite{r71,r72,r73,r74,r75,r76,r77,r78}. The behavior of current in disordered systems has been 
well studied in the literature~\cite{r79,r80,r81,r82,r83,r84,r85,r86,r87,r88,r89,r90,r91}. In general the current gets reduced with disorder
strength~\cite{r92,r93,r94,r95,r96,r97,r98,r99,r100}. But upon some specific arrangement between the site energies and hopping dimerization 
the current can also be increased with disorder strength. The enhancement of current due to the interplay between site energies and hopping
dimerization has been explored in previous studies~\cite{r101,r102}. However, {\em the modulation of current in the presence of NH effects,
particularly its dependence on the correlation between hopping integrals has received far less attention. Furthermore, this model offers a
unique capability to sustain a purely real current up to a certain threshold of the NH parameter, despite the system being NH in nature. 
This distinct feature emphasizes the novel behavior introduced by the NH factor.} To explore different important characteristics of
current in presence of environmental interaction, we take a ring and incorporates physical gain and loss alternatively only at the odd sites 
of the sample where the even lattice sites are kept free.

\vskip 0.2cm
\noindent
\underline{\em Exceptional Points}: Exceptional points (EPs) are non-Hermitian degeneracies where two or more eigenvalues 
and their corresponding eigenvectors coalesce. These singularities occur in a parameter space of non-Hermitian systems and lead to 
unconventional physical phenomena. EPs are exploited in various advanced applications due to their extreme sensitivity to perturbations. 
In photonics, EPs enable ultra-precise sensors and unidirectional light transport. They are also applied in designing robust lasers with 
controlled mode selection, and in non-Hermitian quantum systems for manipulating state dynamics. Their unique topology offers promising 
platforms for enhanced control in next-generation quantum technologies.

In this work, we focus on various aspects of a non-Hermitian (NH) ring system, including the effects of the environmental interaction 
parameter $d$ and hopping dimerization on the band spectrum, transitions between real and complex eigenvalues, exceptional points, circular 
currents, among others. We classify the real and imaginary parts of the energy spectrum into their respective domains and evaluate the 
ground state energy and current by differentiating the ground state energy with respect to the magnetic flux. This investigation sheds 
light on several key features of flux-driven circular currents, and our analysis may help reveal complex behaviors in other loop geometries 
as well.

This manuscript is structured as follows. We begin with an introduction to the study in Section I. Section II presents the theoretical model based on a tight-binding approach and outlines the key formulas used to calculate the current. Section III contains the results and their discussion, and Section IV contains the summary of the main findings.
 
\section{Quantum ring and theoretical description}

In this section, we provide the quantum system, model Hamiltonian, and the description of theoretical framework for the calculation. 
Our chosen system is a one-dimensional ring, which is influenced by an AB flux, denoted as $\phi$. We utilize a tight-binding (TB) prescription to illustrate the system. In this framework, we can easily capture the discrete nature of the lattice and the quantum effects induced by the 
AB flux. The form of the TB Hamiltonian reads as 
\begin{align}
H &= \sum_{j(odd)}(t_{1}e^{i\theta}c_{j}^\dagger c_{j+1} + t_{1}e^{-i\theta}c_{j+1}^\dagger c_{j}) \nonumber\\
& +\sum_{j(even)} (t_{2}e^{i\theta}c_{j}^\dagger c_{j+1}+ t_{2}e^{-i\theta}c_{j+1}^\dagger c_{j}) \nonumber \\
& + \sum_{j} \epsilon_{j} c_{j}^\dagger c_{j},
\end{align}
where $j$ is the site index, and $c_j^\dagger$, $c_j$ are the usual fermionic operators. Two different types of nearest-neighbor hopping (NNH) 
strengths, $t_1$ and $t_2$, are taken into account to reveal the hopping dimerized case. $\theta$ is the phase factor due to the magnetic 
flux $\phi$ (measured in unit of elementary flux-quantum $\phi_0=ch/e$, $c$ is the speed of light, $e$ is the electronic charge, $h$ is the Planck's constant) in the ring and it is expressed as $\theta=2\pi\phi/(N\phi_0)$. The parameter $\epsilon_j$ denotes the on-site energy. For different sites it takes the form:\\
$\bullet$ $\epsilon_{j}= +i d$ for odd $j$ ($=1, 5, 9, \ldots$)\\
$\bullet$ $\epsilon_{j}= -i d$ for odd $j$ ($=3, 7, 11, \ldots$)\\
$\bullet$ $\epsilon_{j}= 0$ for even $j$ ($=2, 4, 6, \ldots$).\\
The factor $d$ is the NH parameter and it measures the strength of environmental interaction with the ring sites.

We want to explore the characteristics of flux-driven circular current~\cite{r84,r85,r86,r87} in the NH ring system. Specifically, 
we focus on how exceptional points and band spectra influence the current. To calculate the current, we use the definition
\begin{align}
    I^{re/im} = -c \frac{\partial{E_{0}^{re/im}}}{\partial{\phi}}
\end{align} 
where $E_0^{re/im}$ is the real or imaginary ground state energy, associated with the electron feeling. Unlike a hermitian system where only real energy eigenvalues are obtained, for a NH system we encounter both real and imaginary eigenvalues, and thus, we need to compute $E_0$ 
in both the real and imaginary spaces. Diagonalizing the TB Hamiltonian matrix (for an $N$-site ring, the dimension of the Hamiltonian 
matrix becomes $N\times N$), we find all the real and imaginary eigenvalues, and then summing over the lowest $N_e$ energy levels ($N_e$ corresponds to the total number of electrons), we compute $E_0$, at absolute zero temperature, in two different sub-spaces. If $E_m$ is the eigenvalue of $m^{th}$ state then the ground state energy is expressed as 
\begin{align}
E_0^{re/im} = \sum_{m=1}^{N_e} E_m^{re/im}.
\end{align}
In our study, we compute currents in all the cases by fixing total number of electrons $N_e$.

\section{Numerical results and discussion}

In this section, we present our numerical findings and provide insights into the circular current in our system, which is influenced by 
environmental interactions. For our analysis under the condition of unequal hopping integrals (i.e., the dimerized scenario), the values of $t_1$ are maintained, unless mentioned otherwise, at $1.5$ and $0.5$, while the hopping $t_2$ is kept at $1$. When $t_1=t_2$, we denote this condition as $t$ and fix its value at $1$. Our investigation is conducted at half-filling, allowing us to effectively study the behavior of the system. Additionally, key parameters are elaborated upon in the captions accompanying each figure, ensuring that readers have a clear understanding of the relevant conditions and context for our findings. Unless specified, the results are computed for $N=16$. All values of the hopping integrals and energy eigenvalues are expressed in the unit of eV while the current is measured in $\mu$A. Our results include different key aspects and here we discuss them one by one as follows.

\subsection{Energy level spectrum}

In Fig.~\ref{fig:f2} we show the variation of different energy levels as a function of magnetic flux $\phi$ for the dimerized-free ring
($t_1=t_2$). The left column shows the results for the real eigenvalues, while the right column is for the imaginary eigenvalues. Three 
different rows are associated with three distinct values of the NH parameter $d$. For $d=0$, the spectrum consists entirely of real 
eigenvalues, which is expected, as in this case the system is hermitian. The energy levels in the real spectrum exhibit finite slopes and demonstrate high degeneracy, with some slopes canceling each other out, which limits the flow of current within the system. Increasing the value of $d$ to $2$ results in a breakdown of the high degeneracy of some energy levels. The slopes of certain levels become steeper, and the mutual cancellation of slopes appears less pronounced, suggesting a potential increase in current compared to the $d=0$ case. When $d=4$, the values of real eigenvalues become extremely small, nearly zero. Consequently, the spectrum is dominated by imaginary eigenvalues, leading to a scenario where the real current approaches zero.

The appearance of no imaginary eigenvalues at $d=0$ suggests vanishing imaginary current. As $d$ increases to $2$, the imaginary spectrum develops steeper slopes that generally do not cancel out across most of the flux regimes. This indicates the presence of a finite amount of imaginary current in this scenario. Additionally, at $d=2$, the imaginary part of the energy spectrum looks similar with the real energy spectrum, leading to an equivalent contribution to the current from both the real and imaginary components. This correspondence between the two spectra results in identical current magnitudes in both the real and imaginary spaces. Such behavior highlights the critical role of the gain-loss parameter $d$ in balancing the real and imaginary currents at a specific value, further emphasizing the unique characteristics of the system. The equalization of energy spectrum in both spaces is a distinctive feature that warrants deeper exploration, as it underscores the interplay between real and imaginary energies at this parameter value. Another significant observation at $d=2$ is the emergence of doubly degenerate eigenvalues at the edges of the energy spectrum. This behavior is not limited to the spectral boundaries; within the central regions of the spectrum, we encounter nearly fourfold degenerate eigenvalues in both the real and imaginary eigenenergy spaces. The appearance of such degenerate states in different parts of the spectrum is a remarkable and critical finding. It suggests a deeper underlying symmetry or structural characteristic in the energy landscape, particularly influenced by the gain-loss parameter $d$. This discovery of degeneracy, both at the spectral edges and within the bulk, provides valuable insights into the intricate behavior of the system and warrants further investigation into its physical implications, particularly in terms of current transport in NH regimes. Now, when $d=4$, the spectrum exhibits a gapped structure with three quasi sub-bands. The quasi-bands near the edges of the spectrum are nearly flat, resulting in very small slopes. For 
the quasi-band near the middle of the spectrum, some slopes cancel each other out, reducing the likelihood of significant current compared to the $d=2$ case.

Figure~\ref{fig:f3} displays the variation of real and imaginary energy eigenvalues of the NH ring, when $d$ is set at $2$, for the two dimerized conditions where the top and bottom rows are for $t_1>t_2$ and $t_1<t_2$, respectively. These plots reveal a notable contrast and switching behavior between the real and imaginary eigenspaces depending on the relationship between the hopping integrals. For $t_1>t_2$, we observe the formation of two distinct sub-bands in the real energy spectrum, while the imaginary energy spectrum exhibits an entirely different profile (top row). This behavior is reversed when the condition on the hopping integrals is switched from $t_1>t_2$ to $t_1<t_2$ (bottom row), resulting in a starkly contrasting structure in both the real and imaginary spectra. Furthermore, a column-wise comparison of the spectra shows that altering the relationship between the hopping integrals significantly changes the behavior of both the real and imaginary energy spaces. This distinct switching and contrasting behavior between the two conditions not only highlights the complex interplay between real and imaginary energy spectra but also reveals an intriguing pattern that underscores the sensitivity of the system to the relative strength of the hopping integrals.  

It is important to note that sorting the real and imaginary eigenvalues separately is essential for computing current by summing energies 
up to specific levels for a given filling factor. This band spectrum may provide intriguing and contrasting non-trivial characteristics in 
real and imaginary currents, in presence of the NH parameter $d$. These issues will be discussed in the appropriate parts of our forthcoming 
discussion.

\subsection{Variation of current with flux}

In Fig.~\ref{fig:f4}, we show the variation of the real and imaginary circular currents as a function of $\phi$. These two current components are obtained by separately calculating the ground state energies in two different energy subs-paces, followed by taking the first-order derivative with respect to the magnetic flux. This approach provides a comprehensive view of the behavior of the system. The figure is 
organized such that the left column depicts the real current, while the right column focuses on the imaginary current, allowing for a 
clear comparison between these two components. Three different rows, from upper to lower, correspond to the cases of $d=0$, $2$, and $4$, respectively. This progression enables us to observe how increasing the gain and loss parameter impacts both the real and imaginary currents at each flux value. The results shown here are worked out for the dimerization-free ring ($t_1 = t_2$). This choice ensures that any changes in the current behavior can be attributed directly to variations in the flux and the parameter $d$, rather than to asymmetry between the hopping integrals.

Focusing on the real current, we observe a gradual and steady increase with the gain and loss parameter $d$. This trend is in line with the behavior seen in the energy band spectrum for real eigenvalues, where the real eigenenergies grow as $d$ increases. However, as $d$ reaches the value of $4$, the real current diminishes to zero. At this point, the real eigenenergies of the system become negligible, essentially vanishing. This phenomenon is clearly represented by the orange curve in the plot, illustrating how the real current is directly tied to the real eigenvalue spectrum of the system and becomes insignificant as the real part of the eigenspectrum fades away at higher values of $d$.

Turning our attention to the right column of the figure, we can see that the imaginary current behaves differently. When $d=0$, the imaginary current is zero, which is expected as the system exhibits a completely real eigenspectrum in this regime. As $d$ increases, the imaginary current begins to rise, reflecting the increasing contribution of imaginary eigenvalues. However, this rise is not indefinite. At $d=4$, we notice a reduction in the imaginary current. This oscillatory pattern in the current is largely driven by the interaction between physical gain and loss, which occurs at alternating odd lattice sites. Interestingly, even though the real energy spectrum becomes almost entirely imaginary at $d=4$, this does not lead to a continuous enhancement of the imaginary current. Instead, the current follows a non-linear behavior due to the complex effects of gain and loss. The parameter $d$ influences the hopping process, and once a certain threshold is surpassed, localization effects may arise. This suggests that while the gain and loss terms contribute to the current, they also introduce a limit beyond which further increment of $d$ does not result in enhanced current flow. Rather, they might lead to the suppression of transport due to the onset of localization effects, which restrict the mobility of particles in the system and thereby reduce the magnitude of the current.

An additional significant observation from Fig.~\ref{fig:f4} is the intersecting point where both the real and imaginary currents approach similar values, particularly around $d = 2$. This is a key result, indicating that at this particular value of the gain and loss parameter $d$, the magnitudes of both the real and imaginary components of the current are equal. This phenomenon is distinctly visible in the figure, highlighting the intersection of the two current profiles, where the contributions from real and imaginary eigenvalues appear to be balanced. The equivalence of the real and imaginary currents at $d=2$ can be further clarified by examining the underlying relationship between the energy spectrum of the system and the applied flux. As the energy bands evolve with flux for this value of $d$, it appears that the interplay between real and imaginary eigenvalues becomes symmetrical, resulting in the matching currents. This behavior suggests that at $d=2$, the system reaches a unique point of balance between the gain and loss effects introduced at alternate lattice sites. Such a scenario is particularly important because it emphasizes the critical role that the parameter $d$ plays in dictating the transport properties. This observation not only contributes to our understanding of the role of $d$ in modulating the current but also hints at potential symmetry or resonance effects that arise at intermediate values of the gain and loss parameter, further enriching the overall picture of transport in NH systems.

If we closely examine Fig.~\ref{fig:f5}, an intriguing pattern emerges. In this figure, we plot the current as a function of flux, where the orange and blue curves represent the cases $t_1>t_2$ and $t_1<t_2$, respectively. At $d=0$, the real current for $t_1>t_2$ exceeds that of $t_1<t_2$, suggesting that the system with stronger hopping integral ($t_1$) displays a more significant transport response. However, once we introduce the NH parameter $d$ and set it to $d=2$, the situation reverses. In this case, the current for $t_1<t_2$ becomes greater than for $t_1>t_2$, highlighting a shift in the transport characteristics driven by the gain and loss parameter. What makes this even more intriguing is that in the imaginary current space, the behavior is inverted at $d=2$. In other words, while the real current for $t_1<t_2$ dominates, the imaginary current shows the opposite trend, with $t_1>t_2$ exhibiting a larger imaginary current than $t_1<t_2$. This inversion points to a non-trivial interaction between the real and imaginary components, suggesting that such type of gain and loss in NH systems can lead to distinct transport signatures depending on the domain being considered. Moreover, in imaginary space this behavior persists even when $d$ is further increased to $d=4$, where the real current has diminished to zero. Even at this value, the imaginary current for $t_1>t_2$ continues to exceed that for $t_1<t_2$, indicating that the switching behavior between different hopping correlations, as modulated by the parameter $d$, is indeed an interesting feature. These recurring transitions between different hopping regimes as $d$ varies are undoubtedly significant findings.

These plots play a pivotal role in demonstrating how the current evolves in response to changes in $d$, offering a clear visualization 
of their interaction and influence on the transport properties of the system. By calculating the current at a fixed flux and analyzing 
its variation with $d$ for both the real and imaginary components, we gain valuable insights into the underlying physical mechanisms. 
This approach allows for a comprehensive understanding of the NH effects on current behavior. The following sub-section will provide 
a detailed analysis and discussion of these significant observations.

\subsection{Variation of ground state energy with flux}

To better understand the behavior of current at exceptional points, we inspect the variation of both the real and imaginary components of the ground state energies with respect to the magnetic flux (left column of Fig.~\ref{fig:f6}), focusing on the values of $d$ near the first exceptional point for a specific case where $t_1=t_2$. For a clearer view of the variation of the slope, we chose a small flux window and
re-plot the ground state energies (right column of Fig.~\ref{fig:f6}). The results indicate that the ground state energies change continuously with flux, marked by sharp slopes. These steep gradients have a pronounced impact on the current at exceptional points, significantly enhancing both the real and imaginary components of the current as $d$ is varied. This observation highlights the critical influence of the slope steepness on current, showing how the parameter $d$ plays a vital role in amplifying both the real and imaginary parts of the current in the vicinity of these exceptional points.
 
\subsection{Variation of energy and typical current with $d$}

In Fig.~\ref{fig:f7}, we present the variation of real and imaginary eigenvalues together with the two current components as a function of
the gain-loss factor $d$, setting the magnetic flux $\phi=0.3$. The three different rows represent the three different conditions of the 
hopping integrals. Upon examining the energy spectrum, we observe that the real eigenvalues bifurcate at lower values of $d$ and approach the zero energy axis at higher $d$. Conversely, the imaginary eigenvalues exhibit the opposite behavior to the real ones. Exceptional points are identified in both the real and imaginary spectra. Given these characteristics of the energy spectrum, it is intriguing to investigate the behavior of the current at these exceptional points on the zero energy axis. We compute the currents at $\phi=0.3$, and show their dependence on the gain-loss parameter $d$. For the real current, we observe that it is lower at the bifurcation points away from the zero energy axis. As the energy spectrum approaches and intersects the zero energy axis, the current gradually increases, exhibiting peaks at four exceptional points, and diminishes when the real energies become negligible and approach to zero.

Conversely, the imaginary current displays an opposite pattern with respect to $d$. It shows peaks at the four exceptional points on the zero energy axis and decreases as the spectrum bifurcates away from this axis, diverging outward. This contrasting behavior of the real and imaginary currents is clearly illustrated in Fig.~\ref{fig:f7}. We can say that the diverging nature of the energy spectrum from the zero energy axis generally leads to a reduction in the current, regardless of the presence of exceptional points away from the zero energy axis. Conversely, the converging tendency of the energy spectrum towards the zero energy level enhances the current. Moreover, the exceptional points lying precisely on the zero energy axis exhibiting maximum sensitivity, significantly enhancing the current and manifesting as peaks at those points.
\begin{figure*}[ht]
\noindent
{\centering\resizebox*{18.0cm}{6.0cm}{\includegraphics{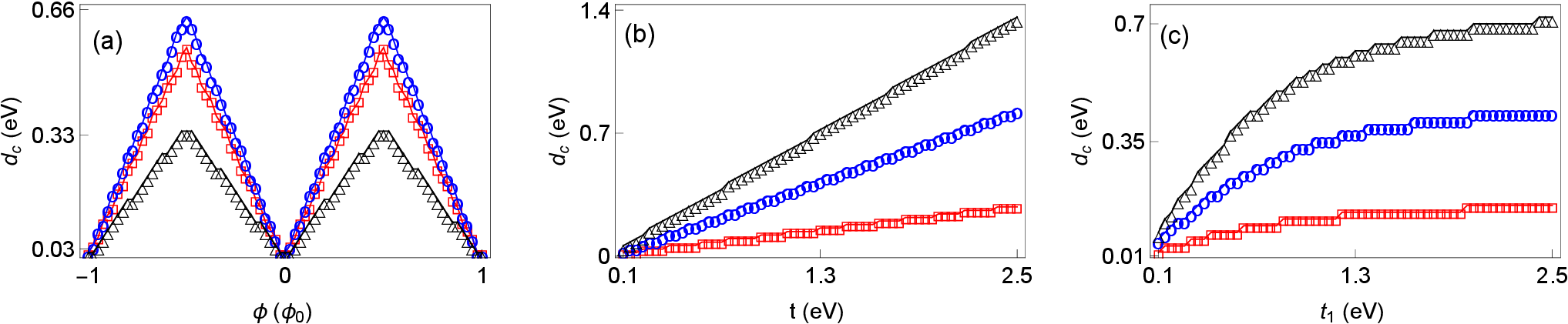}\par}}
\caption{(Color online). Variation of $d_c$ with flux $\phi$ (a), uniform hopping $t\,(t_1=t_2)$ (b), and the hopping ratio $t_1/t_2$ (c). In sub-plot (a), the curves with red squares, blue triangles, and black circles correspond to the scenarios where $t_1=t_2$, $t_1>t_2$, and $t_1<t_2$, respectively. In the other two sub-plots, the curves with red squares, blue circles, and black triangles represent the cases for $\phi=0.1$, $\phi=0.3$, and $\phi=0.5$, respectively.}
\label{fig:f11}
\end{figure*}
When correlations between the hopping amplitudes are introduced ($t_1 \neq t_2$), the energy spectrum experiences a significant shift 
relative to the previous scenario where $t_1=t_2$. This alteration leads to a notable change in the distance between the exceptional points 
in the system, which directly impacts the behavior of the current. Specifically, the spacing between the points where the imaginary current
decreases and the real current increases, as seen in the current spectra with respect to the gain and (or) loss parameter $d$, becomes wider.
This widening occurs for both $t_1 > t_2$ and $t_1 < t_2$ cases, suggesting that the correlation between the hopping terms plays a crucial 
role in altering the response of the system.

In these cases, the real current continues to increase over a larger range of the parameter $d$, while the suppression of the imaginary 
current is also more spread out. This broader separation between the decrement of the imaginary current and the enhancement of the real 
current signifies a more pronounced impact of the exceptional points on the transport properties of the system. The introduction of 
asymmetry between the hopping amplitudes not only shifts the position of the exceptional points but also alters the manner in which the 
currents evolve, giving rise to an extended region where the real current is amplified. By tuning the relationship between $t_1$ and $t_2$,
one can systematically modify the transport properties, thereby offering a potential method for enhancing the real current while 
simultaneously suppressing the imaginary component. This ability to control the interplay between the real and imaginary parts of the 
current highlights the versatility of the system and its sensitivity to changes in hopping correlations.

In Fig.~\ref{fig:f8}, we display the energy eigenvalues and the two current components as a function of $d$, similar to Fig.~\ref{fig:f7}, 
but in this case the magnetic flux is set at the half-flux-quantum ($\phi=0.5$). Notably, this figure reveals a substantial reduction in the number of exceptional points compared to other flux values, which, in turn, leads to fewer pronounced peaks in both the real and imaginary components of the current. A key observation is that, in the hermitian case, the current at $\phi = 0.5$ tends to zero, indicating minimal transport at this flux. However, under NH conditions, we observe a significant non-zero current, despite the reduced number of peaks relative 
to other flux values. This behavior underscores the impact of NH pattern in maintaining and even enhancing current at specific flux values, which would otherwise show minimal current in hermitian systems. Additionally, this phenomenon is consistently observed across all three cases involving different relations between the hopping integrals, although the position and behavior of the exceptional points vary with each case. These results suggest that while the number of exceptional points may be reduced for $\phi = 0.5$, the NH characteristics still facilitate substantial current, highlighting the robustness of current enhancement in such systems despite variations in flux and hopping parameters.

As shown in Fig.~\ref{fig:f9}, we observe that the real component of the current exhibits a noticeable enhancement as the parameter increases.
This upward trend continues until $d$ reaches a specific threshold, which we refer to as the critical value $d_c$. Beyond this point, the
behavior of the current may change, indicating the significance of $d_c$ in governing the response. This critical value marks a key 
transition in the behavior of the current as $d$ is varied, where the imaginary part of the current is zero. Within this range, the system
maintains a completely real behavior, and the increase in $d$ from its initial value leads to a significant amplification of the real current.
This phenomenon is consistently observed across all three configurations of the hopping integrals, though the threshold value $d_c$ varies
between these cases. The presence of such physical mechanisms involving the gain and loss underpins the existence of a purely real energy
spectrum up to this critical value of $d$, allowing for substantial enhancement of the current. This observation highlights the impact of 
the such type of physical gain and loss on its transport properties.

To inspect the behavior of current in a relatively larger ring system in Fig.~\ref{fig:f10}, we present the variation of both the real 
and imaginary components of the current with respect to the gain-loss parameter $d$, under three different conditions involving the 
hopping integrals. The results are shown for two distinct flux values: $\phi = 0.3$ in the upper row and $\phi = 0.5$ in the lower 
row, considering a $200$-site ring. Despite the increase in system size, the 
distinct behavior of real and imaginary currents for the varying hopping conditions remains clearly evident across both flux values. 
This highlights the robustness of the behavior of the current in NH systems, even in larger lattices. Moreover, for $\phi = 0.5$, we 
observe a decrease in the number of peaks in the current, which is in line with the trends previously identified in smaller system sizes. 
This indicates that the response of the system, particularly the reduction in peaks at this specific flux value, scales consistently with 
size. Another noteworthy observation is the substantial magnitude of the current, even for the large site ring. This appreciable value of 
current, despite the larger lattice size, is a crucial finding and demonstrates that the system retains significant transport properties 
as the lattice expands. The figure provides clear visual confirmation of these trends, emphasizing the persistence of these distinctive 
current features in both real and imaginary components for larger quantum systems.
\begin{figure*}[ht]
\noindent
{\centering\resizebox*{16.0cm}{12.0cm}{\includegraphics{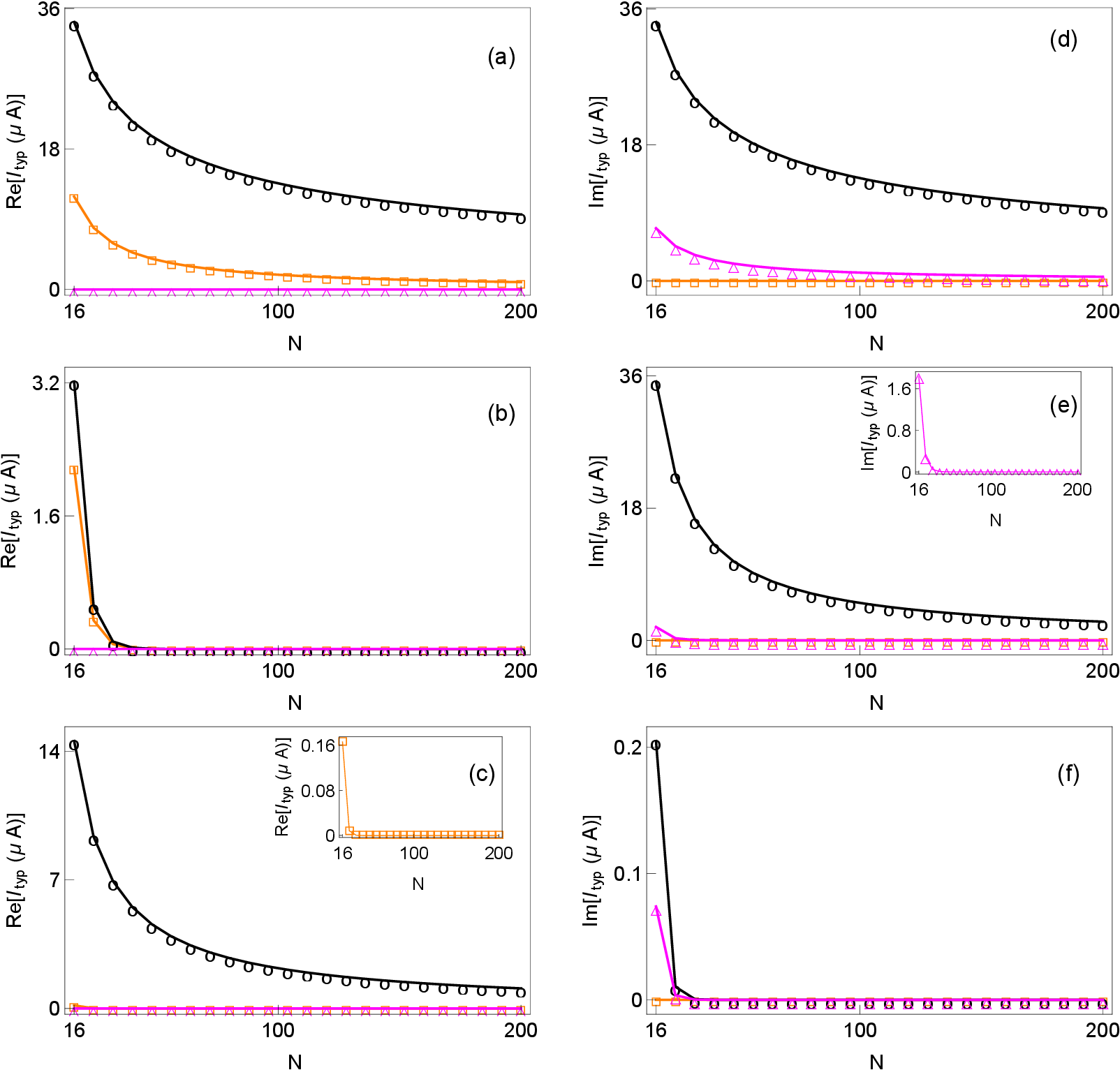}\par}}
\caption{(Color online). Variation of real (left column) and imaginary (right column) current components as function of ring size $N$, by
varying it in a wide range, when the magnetic flux $\phi$ is set at $0.3$. The top, middle, and the bottom rows correspond to $t_1=t_2$,
$t_1>t_2$, and $t_1<t_2$, respectively. In each sub-plot, the orange, black, and the magenta curves with distinct line styles are for 
$d=0$, $2$, and $4$, respectively. To enhance visualization, the insets of sub-plots (c) and (e) show the variation of the real and 
imaginary components of the current, respectively, as a function of $N$.}
\label{fig:f12}
\end{figure*}

\subsection{Variation of $d_c$ with $\phi$ and hopping parameters}

In our analysis, we introduce a parameter, $d_c$, which denotes the critical value of $d$ up to which the imaginary current remains zero. Figure~\ref{fig:f11} illustrates how $d_c$ varies with $\phi$ and the hopping integrals. The sub-plot in Fig.~\ref{fig:f11}(a) specifically depicts the relationship between $d_c$ and the flux for three distinct conditions involving the hopping integrals. The red, green, and blue curves correspond to $t_1 = t_2$, $t_1 > t_2$, and $t_1 < t_2$, respectively. It is evident from the figure that $d_c$ is higher when $t_1 > t_2$ compared to the other cases. Another key observation is that $d_c$ reaches its maximum when $\phi = 0.5$, and then decreases linearly, reaching to zero for integer values of flux. This indicates that for $\phi = 0.5$, the system can sustain higher values of $d$ before the onset of any imaginary current. The interaction between the hopping integrals, the physical gain and loss mechanisms, and the applied flux collectively influence the system, resulting in these observed characteristics. This analysis not only highlights the dependency of $d_c$ on the interplay between these factors but also provides valuable insights into the critical thresholds beyond which the imaginary current emerges in the system.

The occurrence of this specific behavior in $d_c$ suggests that the location of the first exceptional point along the zero-energy axis can be effectively controlled by varying the flux, as well as by adjusting the conditions between the hopping integrals. This indicates that the exceptional point, which marks a critical transition, is not fixed but can be shifted depending on these external parameters. By manipulating the flux and the relationship between $t_1$ and $t_2$, the critical point can be fine-tuned, offering a deeper understanding of how external factors influence the emergence of exceptional points. This tunability is significant as it provides a potential pathway for experimental control over quantum systems exhibiting NH characteristics.

Subplot Fig.~\ref{fig:f11}(b) illustrates how $d_c$ varies as a function of the hopping parameter when $t_1 = t_2 = t$, for three distinct flux values ($\phi = 0.1$, $0.3$, and $0.5$), represented by the red, blue, and black curves, respectively. From this plot, it is evident that $d_c$ enhances linearly with increasing $t$. Furthermore, the value of $d_c$ remains consistently higher for $\phi = 0.5$ compared to the other flux values across all hopping strengths. This behavior suggests that at too low values of hopping, the first exceptional point where the system undergoes a transition into a NH regime begins near zero. As the hopping parameter $t$ increases, the location of the exceptional point shifts to higher values of $d$. In other words, stronger hopping allows the system to support larger values of $d$ before the onset of imaginary current, with the flux value playing a significant role in determining the precise magnitude of $d_c$. The dependence of $d_c$ on both the hopping strength and flux highlights the intricate interplay between these parameters in controlling the transition of the system to the exceptional point.

The variation of $d_c$ with the ratio $t_1/t_2$ is presented in sub-plot Fig.~\ref{fig:f11}(c), where the color-coded curves represent different flux values, as indicated in the earlier sub-plot. This plot reveals an approximately step-like behavior in $d_c$ as $t_1/t_2$ changes. Notably, the position of the first exceptional point shows minimal variation for small changes in $t_1$. The step-like pattern is particularly pronounced for $\phi = 0.1$, especially at higher values of $t_1$. This suggests that under certain flux conditions, the system exhibits a more distinct transition in $d_c$ as the hopping ratio is adjusted, with the influence of $\phi$ being most significant for smaller flux values. This behavior underscores the sensitivity of the exceptional point to the interplay between hopping integrals and flux, revealing that in certain regimes, the system remains robust to small changes in $t_1$, while in others, the response is more pronounced. This step-like nature could have implications for controlling and fine-tuning the transition points in experimental setups.

\subsection{Variation of current with $N$}

Figure~\ref{fig:f12} demonstrates how the system size influences the current specifically in the half-filled scenario. The figure is structured with the real current shown in the left column and the imaginary current in the right column, while the three rows correspond to different configurations of the hopping integrals. Within each graph, the orange, black, and magenta curves represent three different values of $d$.

It is evident from the figure that both the real and imaginary currents decrease as the system size increases, which is indicative of 
the typical behavior of persistent current in larger systems. This decrease in current with increasing system size has been well-documented 
in prior studies and is reaffirmed here. A closer look reveals that as $d$ is increased from $0$ to $2$, there is a noticeable rise in the current. However, when $d$ is increased beyond $2$, the current begins to decrease again, showing a non-linear relationship between the current and $d$. Moreover, an intriguing pattern emerges when comparing the real and imaginary currents under varying conditions of the hopping integrals. For $t_1 > t_2$, the real current is consistently smaller than the imaginary current. On the other hand, when $t_1 < t_2$, this relationship is reversed, with the real current becoming larger than the imaginary current. This inversion in behavior can be explained by the interaction of the gain and loss factors at the odd lattice sites, in conjunction with the specific correlation between the hopping integrals.

Overall, these findings highlight the complex and non-linear nature of the dependence of current on both the system size and the parameter $d$. The findings are consistent with earlier theoretical work on persistent current, while also uncovering new details about the behavior of real and imaginary currents under different conditions. This comprehensive study of how the current varies with system size and other parameters provides a more profound understanding of the underlying physical mechanisms involved.

\subsection{Experimental realization}

An electrical circuit analogue can be devised to mimic the behavior of an SSH ring incorporating complex on-site 
potentials and threaded by an Aharonov-Bohm (AB) flux. In this framework, each lattice site is represented by a distinct node in the 
circuit. The imaginary components of the on-site potential can be engineered using AC voltage sources with specific phase shifts, 
nodes with potentials $\pm i d$ correspond to sources shifted by $\pm90^\circ$, whereas neutral nodes are grounded. Alternatively, 
dissipative and amplifying behaviors can be mimicked by using resistive elements: positive resistance corresponds to energy dissipation 
($-i d$), while negative resistance, realized using active circuitry like negative impedance converters,introduces amplification ($+i d$). 
Hopping between neighboring sites is simulated via capacitors with alternating capacitance values to replicate the staggered SSH coupling. 
The presence of AB flux is modeled through phase factors $e^{\pm i\theta}$ assigned to the coupling elements, where the phase 
$\theta = 2\pi\phi/L\phi_0$ depends on the magnetic flux $\phi$ threading the ring of $L$ sites. These phase effects can be practically 
implemented using circuit components such as gyrators, tunable inductors, or voltage-controlled phase delay elements. The closed-loop 
configuration naturally enforces periodic boundary conditions, allowing for the exploration of quantum interference effects arising 
from synthetic magnetic fields.

\subsection{Implementation of complex potentials at odd sites}

In our non-Hermitian SSH ring model, we strategically implement complex on-site potentials only at the odd-numbered sites, leaving the even-numbered sites neutral (i.e., with zero potential). The odd sites alternate between gain and loss, realized as $+i d$ and $-i d$ respectively. This asymmetric distribution explicitly violates $\mathcal{PT}$ symmetry, as the combined operations of spatial reflection and time reversal do not leave the Hamiltonian invariant. Despite this lack of $\mathcal{PT}$ symmetry, the model exhibits a remarkable transport property: for a finite range of the non-Hermiticity parameter $d$, the system supports purely real-valued current, with the imaginary component vanishing identically. This indicates a delicate balance between energy amplification and dissipation across the lattice that sustains coherent, conservative current flow, even in the presence of non-Hermitian elements.

If, by contrast, one were to impose an alternating gain–loss profile at every site, e.g., $+i d$, $-i d$, $+i d$, $-i d$, and so on, the system would satisfy $\mathcal{PT}$ symmetry. However, in such a configuration, the real and imaginary components of the current would typically exhibit resonances at the same set of $d$ values, complicating the isolation of regimes where one component dominates. In the asymmetric case we consider, the decoupling of peak behaviors allows for selective access to regimes characterized by either purely real or purely imaginary currents.

Such odd site based gain–loss arrangement thus provides greater control over the nature of quantum transport in the system and highlights how non-$\mathcal{PT}$-symmetric configurations can still support physically desirable and tunable phenomena, such as current rectification or unidirectional transparency. 

\section{conclusions}

The primary objective of this study is to investigate the behavior of circular currents and to determine how these currents can be enhanced by carefully tuning parameters that govern physical gain and loss also with the correlations between the hopping integrals. This exploration also takes into account the influence of environmental interactions. By strategically introducing gain and loss components within the ring system, we observe that both the real and imaginary spectra dominate within a particular range of these parameters, leading to the full realization and amplification of the current up to a certain threshold. The emergence of exceptional points, along with the shifting characteristics of the energy band spectrum, plays a pivotal role in significantly amplifying the current. 

Our approach employs a tight-binding model to simulate the system, allowing us to diagonalize the Hamiltonian and extract both the real and imaginary parts of the eigenenergies. The current is then calculated by differentiating the eigenvalues with respect to flux. The analysis reveals that the gain and loss parameters, when finely tuned, can push the system into a regime where the real current is fully enhanced, and the imaginary component either vanishes or complements the real spectrum, depending on the range of parameters. The presence of exceptional points, which mark the transition between different physical regimes, is crucial in this process. These points influence the band structure in such a way that they can substantially increase the current at specific parameter values. The behavior of the band spectrum whether it converges or diverges also directly impacts the magnitude of the current, with converging spectra leading to an enhancement in the current. The key findings of our study can be summarized as follows:\\
$\bullet$ At smaller values of the gain-loss parameter ($d$), the real band spectrum dominates the behavior of the system, gradually diminishing towards zero as $d$ increases for the case where the hopping integrals are equal ($t_1 = t_2$). In contrast, the imaginary band spectrum follows an inverse trend, becoming more prominent as $d$ grows larger.\\
$\bullet$ When $d=2$, both the real and imaginary eigenspectra display similar characteristics, with several eigenvalues exhibiting double and fourfold degeneracies. However, when $d$ increases to $4$, this fourfold degeneracy is broken in imaginary eigenspace, leaving only doubly degenerate eigenvalues.\\
$\bullet$ A key observation is that the behavior of the real and imaginary bands reverses depending on the condition of the hopping integrals ($t_1>t_2$ or $t_1<t_2$) at certain specific values of $d$. This switching behavior offers an interesting insight into how the system transitions between different regimes as $d$ varies.\\
$\bullet$ At $d=2$, for the case of $t_1=t_2$, the real and imaginary currents are identical. However, as $d$ changes, the system displays a switching behavior in the current between the conditions of $t_1>t_2$ and $t_1<t_2$, for both real and imaginary eigenspaces. This indicates a significant correlation between the hopping integrals and the gain-loss parameter.\\
$\bullet$ The sensitivity increases as the real or imaginary energy spectrum converges or diverges near the zero-energy axis. This sensitivity is particularly pronounced when exceptional points lie near the zero-energy axis, where the system becomes especially responsive to changes in parameters.\\
$\bullet$ The number of exceptional points located on the zero-energy axis corresponds to the number of peaks observed in the real and imaginary current spaces. The current reaches its maximum near these exceptional points, where the spectra converge. Interestingly, there is an inverse relationship between the real and imaginary currents: when one peaks, the other dips.\\
$\bullet$ When the hopping integrals differ ($t_1\ne t_2$), the spacing between the drop in imaginary current and the rise in real current becomes larger compared to the case where $t_1=t_2$. This behavior remains consistent even when the system size is increased, suggesting a significant characteristic of the model.\\
$\bullet$ A reduction in the number of exceptional points and corresponding peaks is observed at $\phi=0.5$, while the current increases significantly with the gain-loss parameter in NH systems, compared to Hermitian systems, even at $\phi=0.5$.\\
$\bullet$ Up to a critical value of the gain-loss parameter ($d_c$), the system exhibits a purely real current, with the imaginary current becoming completely zero. Within this limit, the magnitude of the real current can be enhanced by adjusting the same parameter. Importantly, the value of $d_c$ changes based on the relationship between the hopping integrals.\\
$\bullet$ Finally, the variation of $d_c$ as a function of flux, hopping integrals, and the correlation between the hopping integrals reveals several intriguing properties. 

\section*{DATA AVAILABILITY STATEMENT}

All data that support the findings of this study are included within the article.

\section*{DECLARATION}

{\bf Conflict of interest} The authors declare no conflict of interest.

\end{document}